%% file: paper.tex
\documentclass[runningheads]{llncs}

\usepackage[hidelinks]{hyperref}
\usepackage{graphicx}
\usepackage{listings}
\usepackage{color}
\definecolor{mygray}{rgb}{0.98,0.98,0.98}

\input{macros}
\begin{document}
  \raggedbottom
\title{Improving Memory Dependence Prediction with Static Analysis}
%
%
\author{Luke Panayi\inst{1, 2}\thanks{Corresponding author, email: \email{l.panayi21@imperial.ac.uk}} \and
Rohan Gandhi\inst{1, 2} \and
Jim Whittaker\inst{2} \and
Vassilios Chouliaras\inst{2}  \and
Martin Berger\inst{2, 3, 4} \and
Paul Kelly\inst{1, 2}
}
\authorrunning{L.~Panayi et al.}
%
\institute{
  Imperial College London, London, UK \and
  Huawei R\&D UK Ltd \and 
  University of Sussex, Brighton, UK \and 
  Montanarius Ltd, London, UK 
}
\maketitle              
\begin{abstract}
This paper explores the potential of communicating information gained by static analysis from compilers to Out-of-Order (OoO) machines, focusing on the memory dependence predictor (MDP). The MDP enables loads to issue without all in-flight store addresses being known, with minimal memory order violations.
We use LLVM to find loads with no dependencies and label them via their opcode. These labelled loads skip making lookups into the MDP, improving prediction accuracy by reducing false dependencies. We communicate this information in a minimally intrusive way, i.e.~without introducing additional hardware costs or instruction bandwidth, providing these improvements without any additional overhead in the CPU.
We find that across pure C/++ Spec2017 workloads, a significant number of load instructions can skip interacting with the MDP and lead to a performance gain.
These results point to greater possibilities for static analysis as a source of near zero cost performance gains in future CPU designs.

\end{abstract}

\section{Introduction}
Out-of-Order (OoO) execution is a leading source of performance in CPU design by exploiting instruction level parallelism (ILP) in programs to hide memory latencies. Speculative execution, guided primarily by branch prediction, is vital to allowing OoO to scale and exploit as much ILP as possible.
One lesser known component employed in OoO execution is the memory dependence predictor (MDP). The MDP can provide large performance gains by allowing loads to issue as soon as possible (i.e.~without having to wait for all in-flight store addresses to become known) while minimising rollbacks due to memory order violations.
The problem the MDP tackles has many parallels to memory aliasing and memory dependence analysis seen in optimising compilers. In this paper we explore the benefits of using static analysis to guide and supplement existing memory dependence prediction. Previous work has used compiler analysis to provide higher MDP accuracy than the state of the art at the time \cite{replacestoreset}, but this has come at the cost of a hard reliance on the compiler to provide memory dependence prediction. In contrast, we aim to resolve trivial prediction queries ahead of time, while leaving non-trivial queries to be handled by the MDP as normal. This is done in a minimally intrusive fashion, without introducing additional hardware costs or instruction bandwidth, and also means programs that do not use our techniques still run as normal.
We use the well known \emph{Store Sets} algorithm \cite{storeset} as a proof of concept for our idea, as it is a standard reference point for memory dependence prediction benchmarking, and the default algorithm implemented in open source simulators such as a Gem5 \cite{gem5}. We show our methods help to alleviate capacity problems in Store Sets, leading to performance gains.
We achieve this by using LLVM \cite{llvm} to find loads in loops which hold no dependencies with any stores in the same loop. These loads are then labelled and, when issued by the CPU, skip making a lookup into the MDP and always issue as soon as possible. In the event they really are reordered before an aliasing store, the memory order violation is still detected and rolled back as normal, however the labelled loads are still not inserted into the MDP or held back when issued in the future.

\subsection{Memory Dependence Prediction in Out-of-Order Execution}
\label{sec:ooo}
Modern CPU performance is severely limited by memory latencies. To fully utilise a CPU's potential, OoO execution allows the CPU to compute different parts of a program at once depending on what data is already available while it waits for new data to load. This technique is scaled with the help of speculative execution, by, for instance, assuming the result of branch conditions to allow the CPU to compute more of the program in parallel. When a speculation is found to be incorrect, the CPU must rollback uncommitted results and recompute using the correct values. Predictors are used to maximise the rate of accurate speculations and minimise rollbacks.

The best known example of speculative execution is branch prediction, but another important type is memory dependence prediction. Load instructions are often on the critical path for OoO execution, so beginning their execution as soon as possible is important for performance. However, load instructions not only have register dependencies to calculate their address, but memory dependencies too - if a store instruction first writes a value to memory, the CPU must ensure a later load to the same location does not issue first.
To address memory dependencies, address disambiguation in the CPU is done using the load store queue (LSQ), as seen in Section 1 of Figure \ref{fig:lsq}. We show an example of a load dispatching to the relevant components in the issue stage of the OoO pipeline, without speculative execution involved:
\begin{itemize}
  \item A load instruction is inserted into the instruction queue according to its register dependencies
  \item The load is also inserted at the tail of the load queue (LQ)
  \item Once its register dependencies are fulfilled, it issues and searches the store queue (SQ) for in-flight store instructions with matching addresses
  \item If it finds a matching address, the value of the store is forwarded to the load
  \item If it does not find a matching address, the load accesses memory to retrieve its value.
\end{itemize}

\begin{FIGURE}
  \begin{center}
    \includegraphics[width=0.9\textwidth]{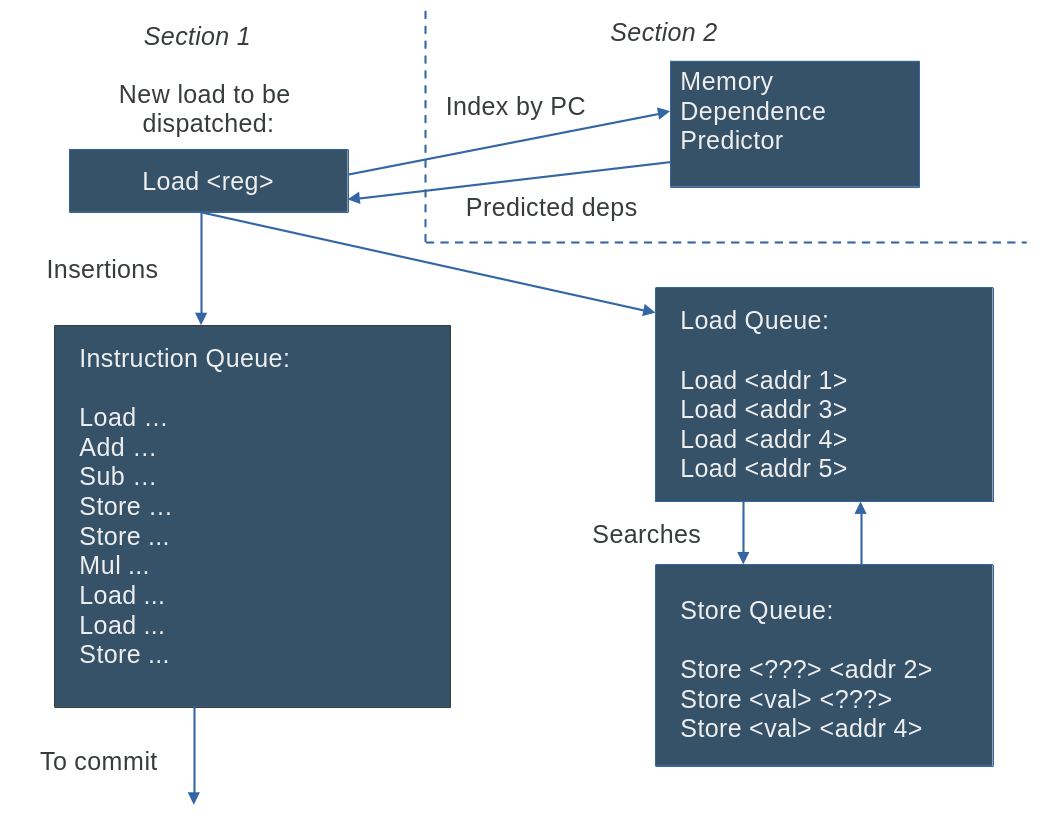}
  \end{center}
\caption{Components involved in issuing load instructions in OoO execution. Section 1 of the figure represents the process without speculative execution, and contains the instruction queue to track register dependencies, the SQ for loads to find forwarding cases, and the LQ for stores to verify proper ordering of loads. Section 2 of the figure introduces speculative execution, and contains the MDP which is PC indexed on load dispatch and returns the PC of stores the load is predicted to be dependent on.}
\label{fig:lsq}
\end{FIGURE}
Now consider the case where an in-flight store does not yet have a computed address. The address of this store could potentially alias with any currently in-flight load, causing a memory dependency. Performance maximising CPUs will therefore speculatively assume in-flight loads do not alias with uncomputed store addresses. When the store eventually executes, it searches the load queue for any younger loads with matching addresses. If it finds one, a memory order violation has occurred and a rollback from the violating load onwards is triggered.
To minimise rollbacks caused by memory order violations, the memory dependence predictor (MDP) tracks which loads have previously caused rollbacks with which stores. When those loads issue in the future, they PC index into the MDP, as seen in Section 2 of Figure \ref{fig:lsq}, and are inserted into the instruction queue according to their register dependencies as well as their predicted memory dependencies.
\subsection{"Predict No Dependency" Load Labels}
A key observation is that process shown in Figure \ref{fig:lsq} happens indiscriminately for every load on every issue - even for loads which could trivially never have dependencies, like accesses to read-only data. If we are confident a load reads memory that is not written to by any in-flight store (i.e., if the OoO window will not contain dependencies for a load), it should be safe for the load to skip making a lookup into the MDP. If we happened to be wrong, and a dependency really did exist, this would still be caught and handled appropriately at the commit stage - all that would change is the speculative decision, not the result. Skipping MDP lookups would eliminate the possibility of false dependencies returned by the predictor (either by index collisions or a real dependency not continuing to hold), and could also yield a power saving due to having fewer cycles where the predictor needs to be accessed. Listing \ref{fig:pnd-example} shows an example of C code where the load on the array $b$ would be a good candidate to label to bypass the predictor. We call these loads "predict no dependency" (PND) loads.
\begin{lstlisting}[language=C, caption=C code that demonstrates an example where a "PND" label could apply. The load on array $b$ has no dependencies in the loop\, and we know $a$ and $b$ will not alias due to the $restrict$ qualifiers., label=fig:pnd-example]
  void PNDExample(int *restrict a, int *restrict b,
                  int n){
    for (int i=0; i < n; i++){
      a[i] += b[i];
    }
  }
\end{lstlisting}
Note that a dependency with the memory location accessed by $b$ may really exist when executed by the CPU - for instance, if the location is written to just before the function call and the OoO window is large enough to contain both loads and stores at once. We discuss such scenarios in Section \ref{sec:limitations}.
\subsection{Contributions}
This paper demonstrates a use of static analysis to find loads unlikely to have dependencies and prevent them from interacting with the MDP. We generate AArch64 binaries with labelled loads, then simulate their effects in Gem5 on a subset of Spec2017. Specifically, we:
\begin{itemize}
        \item[$\bullet$] Implement a LLVM IR analysis pass to iterate over load instructions in loops and determine if they are candidates for PND labels, communicating labels without introducing any additional hardware overhead or instruction bandwidth.
        \item[$\bullet$] Achieve a notable reduction in MDP lookups per kilo-instruction across pure C/++ Spec2017 workloads, averaging 13\% with a peak of 62\%. This demonstrates that a significant amount of program behaviour can be understood ahead of time through static analysis rather than handled by the CPU in certain workloads. This may also make CPUs more power efficient.
        \item[$\bullet$] Achieve a geometric mean performance gain of up to 0.7\% across Spec2017, with individual speedups as high as 3.8\%.
\end{itemize}

\section{Finding PND Labels with LLVM}
\label{sec:llvm}
We use an IR level pass in LLVM to determine load labels. This has the benefit of giving us access to stronger analysis techniques than what is available at the machine code level (such as SCEV \cite{scev} or MemorySSA \cite{memoryssa}). However this also poses a challenge of how to "track" labelled IR loads down to real load instructions in emitted executables. In this section we overview the algorithm to find loads that should be labelled, the current limitations of the analysis, how we track IR values down to machine code emission, and how labels are then communicated to the CPU.
\subsection{Analysis Algorithm}
\label{sec:alg}
Our LLVM pass has the scope of loops and loop nests. For every load in a loop nest, we iterate over every store and every call across the nest. For each load-store/load-call pair, we query LLVM's standard dependency analysis \cite{llvmdep} if a dependency could exist between them. This analysis automatically makes an aliasing query (or a mod/ref query for calls) which determines if instructions could ever access the same memory location. If they cannot, we know no dependency can exist. If they could, LLVM employs further analyses to consider factors such as loop stride patterns to try and prove the dependency can't exist. Loads which are found to never have a potential dependency with any stores or calls in the loop nest are marked as PND loads.
We note that currently our pass uses 'off the shelf' analysis to make queries. We do not employ novel techniques to determine if a load could be labelled as PND, and only use standard LLVM alias and dependency analysis. This means we can expect the compilation overhead of our analysis to be minimal, as the required analyses will already have been run for use in other passes, and with proper pass management by LLVM the instructions we iterate over will already have been loaded into cache. However, we also discuss possible novel techniques that could be used in the future in Section \ref{sec:futurework}, and this may make our analysis more demanding.

\subsection{Analysis Limitations}
\label{sec:limitations}
As stated, our pass has the scope of loop nests. However, OoO execution has no conception of loop or even function boundaries; it is perfectly possible that a load which holds no dependencies within a loop may depend on a store before the loop - either in the same function or in the caller of the current function.
Therefore, loads which are labelled PND are still searched against by committing stores, and so if memory order violations do occur they will be found and rolled back. This means we do not necessarily need to be exact in our analysis of finding labelled loads.
We can safely assume that dependencies that cause a single violation (or as we'll see in Section \ref{sec:cpu}, occasionally with a sufficiently long period) are not a problem. Even in the base case without any PND labels, these violations will still occur as the predictor will not have previously seen them and so cannot possibly prevent them. So we can safely gain the benefits of labelling a load as long as we expect it to cause a violation infrequently enough.
However, a corner case exists in which labelling a load can cause repeated violations and cause slowdowns. This can occur when two loops are sufficiently close to each other in a program, and can exist in the same OoO window at once. An example is given in Listing \ref{fig:limitations}. If the load on $b$ in the second loop is labelled as PND, it could potentially be repeatedly reordered above the stores on $b$ in the first loop as the OoO window slides across both.
\begin{lstlisting}[language=C, caption=C code demonstrating a corner case which could potentially cause repeated memory order violations with our current labelling pass. To keep the example concise assume no compiler optimisations are applied., label=fig:limitations]
  void PNDLimitation(int *restrict a, int *restrict b,
                     int n){
    for (int i=0; i < n; i++){
      b[i] = i;
    }
    for (int i=0; i < n; i++){
      a[i] += b[i];
    }
  }
\end{lstlisting}
The current analysis makes no effort to catch these cases, but we will see in Section \ref{sec:evaluation} that a naive approach to labelling loads is currently sufficient. In Section \ref{sec:futurework} we discuss further ways of addressing the problem, should it ever cause slowdowns on different workloads, or as we find more loads to label with stronger analysis.

\subsection{Compiler to CPU Communication}\label{sec:communication}
To achieve our goal of being minimally intrusive, we implement load labels by introducing a new set of load opcodes into the AArch64 ISA. These load opcodes work exactly like regular loads, but behave differently when issued by the CPU. They skip MDP lookups to always issue as soon as possible, and do not write an entry into the MDP in the event they cause a violation. During code emission, we find loads that have been determined to be suitable to label and change their opcodes to the corresponding labelled version.
In order to track which IR loads we want to emit with labelled opcodes in LLVM, we extend the $AAMDNodes$ metadata as this is the only type of metadata in LLVM which is preserved when lowering IR to machine code. We run a simple second pass in the LLVM AArch64 backend which iterates over machine instruction loads looking for this metadata, and changes the opcode to the corresponding labelled version before code emission.

\section{Simulation and Workflow}
\label{sec:gem5}
In this section we overview how load labels are implemented in the Gem5 simulator, the parameters of different CPU sizes we use in simulation, and how we use SimPoints \cite{simpoints} to speed up benchmarking.

\subsection{Experimental Design}
Our hypothesis is that even lightweight analysis can lead to a significant reduction in MDP lookups while overall improving performance. To test this we use Gem5 to simulate benchmarks compiled with and without our labels, then compare simulation counters between the two runs. We are primarily interested in how effectively our pass labels frequently executed loads (i.e.~the "code coverage" we achieve), and differences in performance, measured by Cycles per Instruction (CPI).
We  run each benchmark on three different CPU size configurations, to study how our labels behave as parameters in OoO execution vary. For instance, we may expect a larger window size to lead to more violations than smaller ones, due to capturing more memory dependencies and increasing the probability of running into the corner cases described in Section \ref{sec:limitations}.
\subsection{Simulating PND Labels in Gem5}
Once we have compiled labelled binaries with LLVM, we run them under Gem5 to measure the effects of our changes. We implement labels by adding our new opcodes to the AArch64 Gem5 frontend, then implement their semantics by adding a flag to the instruction object class. This flag is set when one of our new opcodes is decoded, and can then be checked against by the relevant components of OoO execution.
We also add a counter to measure how many times a lookup is made into the MDP during the simulation. This is especially important as it indicates the coverage our LLVM pass was able to achieve. By comparing the number of lookups made by the labelled and unlabelled programs, we can see how many issued loads were labelled loads, and have an idea of if we're making any effective change at all.

\subsection{Gem5 CPU Configuration and the Store Sets Predictor}
\label{sec:cpu}
We cover briefly the different size configurations we test and how the MDP in Gem5 relates to them. We three CPU configurations that  resemble a modern phone (\emph{small}), modern workstation (\emph{large}), and a next generation workstation (\emph{extra large}). All configurations are found in Table \ref{table:cpu}.

To understand configurations related to the MDP we briefly overview the Gem5 MDP algorithm. Gem5 implements the Store Sets predictor \cite{storeset}, which consists of two tables, the SSIT (Store Set ID Table) and LFST (Last Fetched Store Table), whose sizes must both be powers of two. The SSIT is PC indexed by instruction program counters, hence index collisions lead to false dependencies.  Consequently, increasing SSIT size reduces the chance of false dependencies. Store Sets also has a clear period value, which is the number of memory operations issued before clearing all entries in the predictor. This prevents the tables from becoming saturated. We scale the period parameter with the table sizes as well, so, as the predictor grows larger, it also resets less often.
The default table sizes in Gem5 are both 1024 entries, which is unrealistically large for modern machines. We scale this down for all configurations, and scale the clear period length by using the ratio between the SSIT size and number of memory operations. In Gem5 the default clear period is 249856 memory operations, so divided by 1024 this gives a ratio of 244. For new table sizes, we find the new clear period by multiplying the table size with this ratio.

{{\small
\begin{center}
  \label{table:cpu}
  \begin{tabular}{ |p{3.5cm}|p{2cm}|p{2cm}|p{2cm}| }
    \hline
    Component & Small & Large & Extra Large \\
    \hline
    Pipeline Width & 8 & 12 & 12 \\
    Inst Queue Entries & 64 & 192 & 384 \\
    ROB Entries & 192 & 576 & 1024 \\
    LSQ Entries & 32 & 96 & 192 \\
    SSIT/LFST Entries & 32 & 128 & 256 \\
    Clear Period Length & 7808 & 31232 & 62464 \\
    L1i/d Cache & 32KiB & 128KiB & 256KiB \\
    L2 Cache & 512KiB & 2MB & 4MB \\
   \hline
  \end{tabular}
\end{center}
}}

\subsection{Simulation Workflow}
\label{sec:workflow}
As detailed simulation in Gem5 is exceptionally slow, we use the SimPoints methodology \cite{simpoints} to generate snapshots of representative regions in our benchmarks and only simulate these on the slow, detailed model (\emph{O3 CPU}). As explained, we compile two sets of binaries for each benchmark, one with labelled opcodes and one without. Binaries without  labelled opcodes can run natively. As both sets of binaries are identical outside of labels, we use Valgrind \cite{valgrind} to generate the basic block vectors for use by SimPoints, speeding up the generation of checkpoints significantly.
We then use the generated SimPoints to do a single full Gem5 run using the fast, undetailed model (\emph{Atomic CPU}), which generates the checkpoints. We simulate with the labelled binaries here, so that the generated checkpoints include the labelled opcodes in the snapshot of the memory state. This yields one set of checkpoints which can be used to simulate both labelled and unlabelled binaries; when we want to simulate the unlabelled case, we disable checks for PND flags in Gem5. This means behaviour is as if the binaries were unlabelled.

\section{Evaluation}
\label{sec:evaluation}

We simulate the pure C/++ subset of Spec2017 in Gem5 with labels enabled and disabled to compare CPI differences and measure the percent reduction in MDP lookups made over the course of execution. We have excluded \texttt{638.imagick\_s} due to being unable to generate checkpoints in time.
The main barrier to including more benchmarks was the maturity of the LLVM Fortran frontend, $flang$. The vast majority of benchmarks including Fortran code currently fail to compile \cite{flang}. Furthermore, of the few Fortran benchmarks which do compile, IR emitted with $flang$ makes heavy use of MLIR which our LLVM IR based analysis cannot readily analyse without further work.
As outlined in Section \ref{sec:cpu} we run benchmarks on three CPU configurations of varying sizes. We run checkpoints generated using SimPoints (discussed in Section \ref{sec:workflow}) with an instruction interval of 100 million instructions and a warm-up period of 10 million instructions.
Our pass is implemented in LLVM 16 and we simulate with Gem5 version 22.0.0.1.

\begin{FIGURE}
  \begin{center}
    \includegraphics[width=\textwidth]{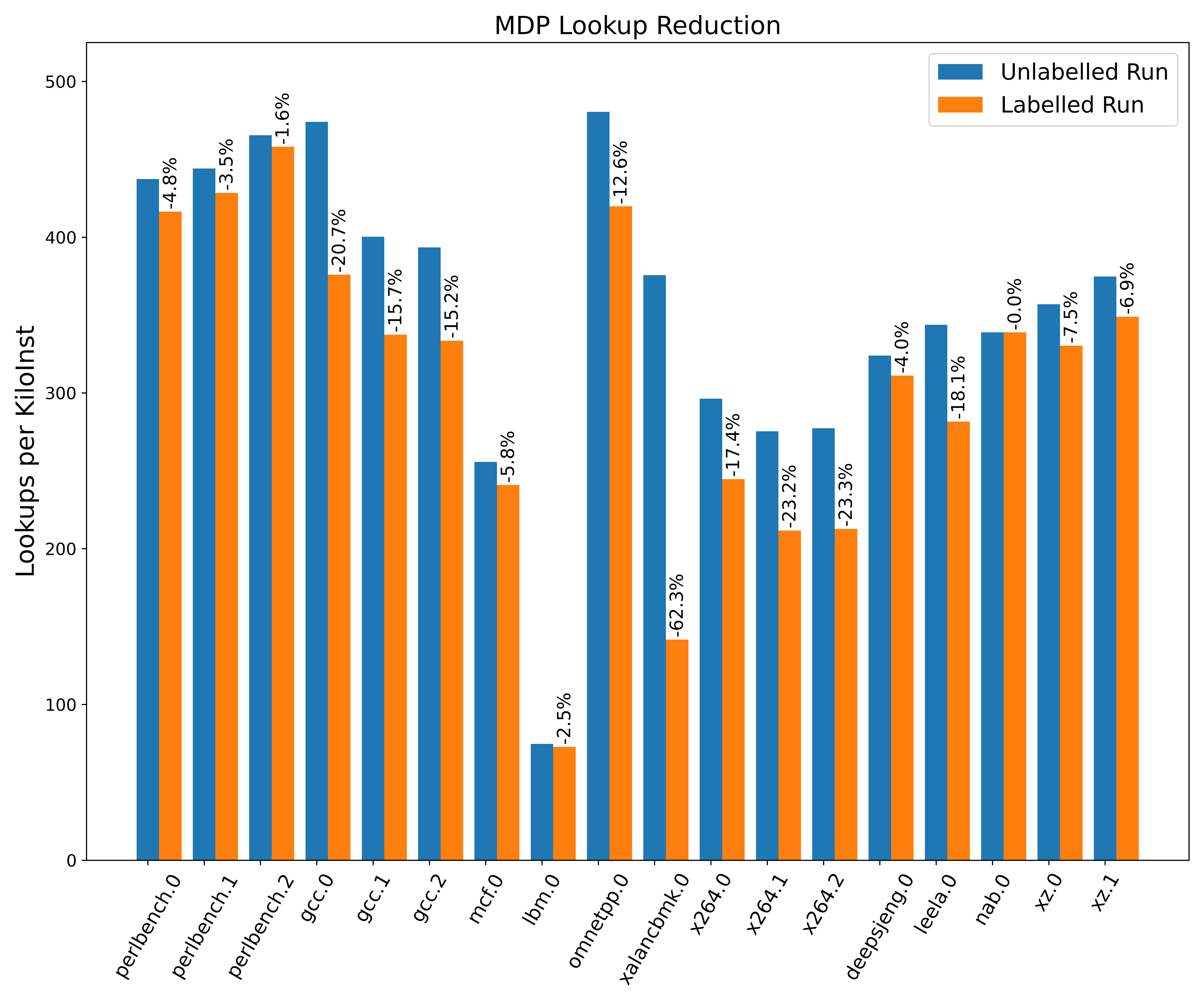}
  \end{center}
  \vspace{-8mm}
\caption{MDP look-ups per kilo-instruction between labelled and unlabelled benchmarks. Lower is better, values are near equal on all three CPU size configurations.}
\label{fig:lookups}
\end{FIGURE}

\subsection{Coverage}
We begin by looking at the code coverage we achieve, which expresses the significance of the loads our pass is able to label. This is more meaningful than the number of loads labelled, as captures how often individual loads are issued over the program. We measure this through the lookups made into the MDP per kilo-instruction, and the percent reduction achieved in labelled runs over unlabelled runs.
As seen in Figure \ref{fig:lookups}, we achieve a mean lookup reduction of 13\% across all benchmarks, but the standard deviation is high. In many cases we struggle to label many loads at all, but in others we can see significant reductions. As we will discuss in \ref{sec:futurework} our analysis can still be pushed further, so we feel these results are already very promising.

\subsection{CPI Over CPU Sizes}
\label{sec:cpi}
 Figure \ref{fig:cpi} shows changes in Cycles per Instruction (CPI) between labelled and unlabelled binaries. A lower CPI means higher performance.
\begin{FIGURE}
  \begin{center}
    \includegraphics[width=\textwidth]{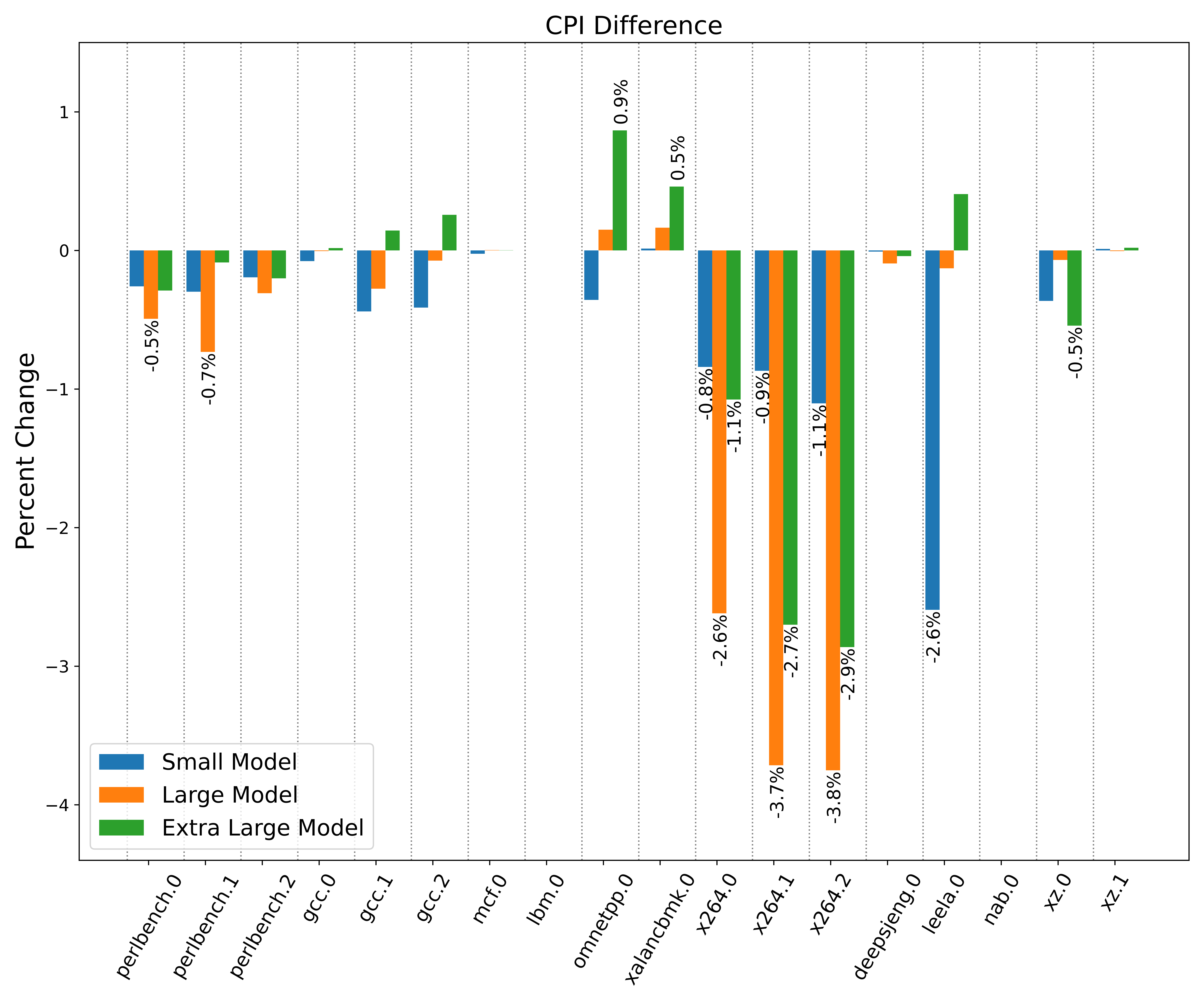}
  \end{center}
  \vspace{-6mm}  
\caption{CPI percent change between labelled and unlabelled binaries. Changes less than 0.5\% in magnitude are unlabelled. Lower is better.}
\label{fig:cpi}
\end{FIGURE}
Overall, we are able to achieve our reduction in lookups without any significant slowdowns. We also achieve notable performance gains in \texttt{625.x264\_s} and \texttt{641.leela\_s} for the small model. One might wonder why, despite having the largest reduction in lookups by far, \texttt{623.xalancbmk\_s} sees no change in performance. We posit this is due to the memory dependence behaviour of the benchmark not being very complex to begin with, meaning there isn't much opportunity for us to beat the predictor. However, it should still be seen as a gain that lookups can be reduced so dramatically without harming performance.

\subsection{Discussion}
\label{sec:discussion}
Analysing the source of performance differences in \texttt{625.x264\_s} and \texttt{641.leela\_s} we found that, as expected, the size of the MDP (SSIT and LFST entries) plays the dominant role most of the time. For instance, \texttt{641.leela\_s}'s performance goes from a strong gain to a small loss as soon as the CPU is scaled past the small model. We find the two larger models regain this performance if their MDP sizes are reduced in size to that of the small model (32 SSIT/LFST entries).
We determine that the performance gains are due to the problem of false dependencies suffered by Store Sets. Because two loads can index to the same SSIT entry, if a memory independent load indexes to an existing entry it will schedule according to a dependency that doesn't really exist. Labelled loads avoid this possibility by never making a MDP lookup, and so the number of false dependencies are reduced. This effect is greatest on smaller table sizes where index collisions are more likely, and so we see as the MDP size grows these benefits are reduced. We confirm this by extending Gem5 to measure the number of index collisions that occur in the MDP, and find the value is significantly reduced on labelled runs.
We determine that the slight performance losses in several benchmarks in the largest model, most notably \texttt{641.omnetpp\_s}, are due to an increase in memory order violations. As the MDP grows, the number of false dependencies in the unlabelled run falls, and as the instruction window grows (the combination of the instruction queue, ROB and LSQ) the number of violations in the labelled run rises. This suggests in these benchmarks we run into the corner case described in Section \ref{sec:limitations} more frequently. However even with the size of a next generation (extra-large) model, this still does not lead to a very significant loss.
This relation between smaller and larger CPU models does not entirely account for what is seen with \texttt{625.x264\_s} however. The large model actually sees larger gains in performance than the small model, despite having a larger MDP and instruction window. We were able to determine the larger instruction window was actually the source of performance gains, rather than working against performance like for instance in \texttt{641.leela\_s}. This implies as the instruction window scales it captures new behaviour which the base Store Sets algorithm does not handle well.
These additional gains are then lost again in the extra-large model, however in this case we isolate the cause as once again just the MDP size. This implies whatever additional behaviour that begins to be captured in the large model is still related to index collisions in some way, however we are currently unable to establish a full explanation of the interaction at play here.
Overall we can determine our benefits over base Store Sets come from a reduction in false dependencies leading from index collisions in small predictors.

\section{Threats To Validity}
We discuss briefly possible threats to the validity or usefulness of the results we present here.
As we do not yet fully understand the source of performance gains in \texttt{625.x264\_s} mentioned in \ref{sec:discussion}, it is possible that the benefits are a result of particular "emergent behaviour" in Gem5 and therefore may not generalise outside this benchmark. Further work is needed to uncover the exact behaviour leading to these performance gains in \texttt{625.x264\_s} and determine if it is reasonable to expect that it will occur in other general purpose workloads.
We acknowledge that our selection of benchmarks are only a subset of Spec2017 and may not be fully representative. As we see large variation in results between benchmarks, it is important that we consider as many kinds of different workloads as possible going forward.
Lastly it should be understood that our CPU model configurations in Section \ref{sec:cpu} are naively scaled, using the default Gem5 configuration as a basis and loosely guided by the CPU models used in the Store Vectors paper \cite{storevectors}. As such our results could differ on more optimised and realistic configurations.

\section{Related Work}
\label{sec:relatedwork}
Existing work on compiler-to-CPU communication for memory disambiguation specifically is sparse but does exist. We overview the most relevant works here and explain how our work is distinct to or extends upon them.
Briefly covering existing MDP algorithms, Store Sets \cite{storeset} is the algorithm implemented in default Gem5 and what this project works with. Store Sets is a widely referenced work in memory dependence prediction, acting as the litmus against which later MDP algorithms have been compared \cite{litsurvey}.
Notable MDP algorithms since Store Sets includes Store Vectors \cite{storevectors} and MDP-TAGE \cite{mdp-tage}. Recently the state of the art has been pushed by the PHAST predictor \cite{phast}, which tracks dependent loads and stores and the control flow path between them.
The most immediately relevant work to this project is \cite{cooperative}. This work proposes the use of binary analysis to label loads which make read only accesses and prevent them from being inserted into the LSQ, in an attempt to improve scalability. The most immediate difference between this work and ours is its focus on the LSQ, with no consideration for the MDP. This means it only sees hardware scalability benefits, rather than direct performance gains like we do. Another difference in our work is moving from binary analysis to a higher level IR in LLVM, allowing the for much stronger analysis, which we intend to extend use of in the future. Lastly \cite{cooperative} increases instruction bandwidth demand with marker instructions before loops, and introduces additional state into the LSQ, so it does not share our principle of being "minimally intrusive".
Work that uses LLVM for similar purposes to ours is \cite{storequeueskipping}. This uses LLVM's alias analysis to insert marker instructions which label how many positions in the LSQ a load instruction can safely skip when making a forwarding search, aiding scalability.
Lastly \cite{replacestoreset} attempts to replace the MDP altogether using profile guided analysis. By marking loads with an index into the LSQ for stores they are expected to be dependent on, they are able to replace the MDP with a very minimal on-chip buffer. When the profile behaviour is accurate, this method outperforms Store Sets for any realistic SSIT size. This reinforces our interest in compiler analysis as a high quality source of information for speculative prediction, however in this case the predictor is replaced altogether. We believe a more reasonable approach is to work with predictors, solving predictable queries ahead of time and allowing the hardware to only focus on hard to statically predict queries.

\section{Conclusions}
\label{sec:conclusion}
We have presented promising results of a new method for carrying out memory dependence prediction in OoO execution. We have shown that simple static analysis can deliver significant reductions in the rate of MDP lookups on select benchmarks, and that this can lead to worthwhile performance improvements especially on CPUs with smaller MDP sizes.
As we push to generalise these results to more benchmarks, we expect to see more performance gains in different benchmarks and possibly further gains on benchmarks where benefits already exist. These could be taken as they are - again, without the need for additional hardware overhead - or potentially justify MDPs with smaller sizes or lower complexity while maintaining near equal levels of performance.

\section{Future Work}
\label{sec:futurework}
\subsubsection{Finding more labels}
An important next step is improving the strength of the LLVM analysis pass, for which there are several promising paths ahead.
Most immediately is inter-procedural analysis. We found that in many cases the percent lookup reduction could be as much as doubled if call instructions were ignored when determining a load label, with no penalty to performance in almost all cases. However, the few places where performance was degraded were too significant to justify ignoring calls everywhere. If we could employ further inter-procedural analysis in addition to mod/ref information, we may be able to enjoy the best of both worlds.
One technique we could introduce is exploiting loop versioning in LLVM, which clones loops into sequential and vectorised versions based on a runtime alias check. We have looked into tracking these versions in loop nests and could prevent comparisons between loads and stores across different versions, as these currently create false dependencies.
Another technique is using LLVM's loop access analysis used in auto-vectorisation to find when a dependency is a "forward" dependency (i.e.~an anti-dependency), which we can safely ignore when determining a load label. This is because in anti-dependencies, the store comes after the load, and so in OoO execution the load can never be reordered "past" the store.
We also want to investigate stack spills as a type of dependency always handled by store forwarding, and so can also safely avoid the MDP. Lastly, if we extend to domain specific contexts, we could make use of the MLIR Affine dialect \cite{affine} and the stronger dependency analysis that comes with it.
\subsubsection{Reducing Additional Violations}
It is possible that, as the strength of our pass increases and finds more loads to label, we run into more cases in which we significantly increase the number of violations. In this case we may want to strengthen our analysis pass to avoid labelling loads that cause repeated violations as described in Section \ref{sec:limitations}. We may also consider a hardware solution like that proposed in \cite{cooperative}, sacrificing our minimally intrusive approach but potentially leading to the highest performance of all options (as we would aim to have both a large number of labels and no increase in violations).
\subsubsection{Modern MDP Algorithms}
While we have used Store Sets as a proof of concept, there are interesting directions to take our research with regards to more modern MDP algorithms such as Store Vectors \cite{storevectors} and PHAST \cite{phast}.
We note that our performance gains are dependent on a reduction in false dependencies due to index collisions, but this issue can also be tackled through entry tagging - i.e., storing additional bits of the instruction PC to be compared against on lookup. An example of this is seen in Store Vectors. If we ran our labelled binaries on such a predictor instead we may not expect to see the same benefits in performance. Instead, the use of labelled loads could aim to achieve near equal performance to a tagged predictor without incurring the hardware cost of tagging, and the implementation considerations that come with it (for instance, decisions to be made when two true dependencies index to the same entry).
The PHAST predictor achieves very high accuracy but at the cost of a more elaborate indexing function, hashing the PC and branch history information together to create the index. This means avoiding lookups into the MDP has greater potential for power savings. PHAST's elaborate indexing also means the lookup can take several cycles to complete, and so there may be very particular cases in which avoiding the lookup can allow a load to issue faster than it would otherwise. Finally, PHAST uses confidence counters for each entry, whose precision can be reduced by index collisions from unrelated loads. Minor performance gains may be available in reducing these collisions.

\subsection*{Acknowledgments}
We gratefully acknowledge funding for this work via the EPSRC ICASE scheme (voucher number 210197) with support from Huawei Research UK.

\addcontentsline{toc}{chapter}{Bibliography}
\bibliographystyle{unsrt}
\bibliography{bibtex}

\end{document}

%% file: macros.tex
\usepackage{amsmath,amssymb}
\usepackage{ifthen}
\newboolean{showcomments}
\setboolean{showcomments}{true}
\ifthenelse{\boolean{showcomments}}
  {\newcommand{\mynote}[2]{
    \fbox{\bfseries\sffamily\scriptsize#1}
    {\small$\blacktriangleright$\textsf{\emph{#2}}$\blacktriangleleft$}
   }
  }
  {\newcommand{\mynote}[2]{}
  }

\newenvironment{FIGURE}{\begin{figure}[t] \centering}{\vspace{-5mm}\end{figure}}